\documentclass{PoS}

\usepackage{graphicx}
\usepackage{amsmath}
\usepackage{mathrsfs}
\usepackage{nicefrac}
\def\beq{\begin{equation}}
\def\eeq{\end{equation}}
\def\bea{\begin{eqnarray}}
\def\eea{\end{eqnarray}}
\def\beqa{\begin{equation}\begin{array}{l}}
\def\eeqa{\end{array}\end{equation}}
\def\eqlab#1{\label{eq:#1}}
\def\figlab#1{\label{fig:#1}}

\def\eref#1{(\ref{eq:#1})}
\def\Eqref#1{Eq.~(\ref{eq:#1})}

\def\Figref#1{Fig.~\ref{fig:#1}}

\def\half{\mbox{\small{$\frac{1}{2}$}}}

\def\quarter{\mbox{\small{$\frac{1}{4}$}}}

\def\barr{\left(\begin{array}{c}}
\def\earr{\end{array}\right)}
\def\bmat{\left(\begin{array}{cc}}
\def\emat{\end{array}\right)}
\def\al{\alpha}
\def\ga{\gamma} 
 
  \def\eps{\epsilon}

\def\si{\sigma} 
\def\th{\theta}  
\def\w{\omega}

\def\pa{\partial}

\def\pa{\partial}

\def\nn{\nonumber}

\def\lag{{\mathcal L}}

\def\3d{3-D}

\def\ol#1{\overline{#1}}

\def\bE{{\bf E}}
\def\bB{{\bf B}}

\def\cE{\mathscr{E}}

\def\bq{\mathbf{q}}

\def\be{\boldsymbol{\epsilon}}
\def\bsig{\boldsymbol{\sigma}}

\title{Proton polarizabilities from polarized Compton scattering: low-energy expansion}

\ShortTitle{Proton polarizabilities from polarized Compton scattering: low-energy expansion }

\author{Nadiia Krupina \\
        Institut f\"ur Kernphysik, Johannes Gutenberg--Universit\"at Mainz, 55128 Mainz, Germany\\
        E-mail: \email{krupina@uni-mainz.de}}

\abstract{
We reexamine the low-energy expansion of polarized Compton scattering off the proton and show that the leading non-Born contribution to the beam asymmetry of low-energy Compton scattering is given by the magnetic polarizability alone, the electric polarizability cancels out.  Based on this fact we propose to determine the magnetic dipole polarizability of the proton from the beam asymmetry. We also present the low-energy expansion of doubly-polarized observables, from which the spin polarizabilities can be extracted.
}

\FullConference{52 International Winter Meeting on Nuclear Physics\\
                 27 - 31 January 2014\\
                 Bormio, Italy}

\begin{document}

Studies of nucleon polarizabilities have recently intensified
fueled by theoretical advances based on chiral perturbation theory
and the current experimental programs at MAMI, HIGS and 
CEBAF facilities, see Refs.~\cite{Griesshammer:2012we, Holstein:2013kia} for fresh reviews. As a result, the Particle Data Group
(PDG)~\cite{Beringer:1900zz}
has recently updated its summary of the dipole electric and magnetic
polarizabilities of the proton, yielding \cite{PDG:2013}:
\begin{subequations}
\bea
\alpha_{E1}^{(p)} &=& (12.0\pm 0.6)\times 10^{-4}\,\mbox{fm$^3$}, \\
\beta_{M1}^{(p)} &=& (2.5\pm 0.4)\times 10^{-4}\,\mbox{fm$^3$}.
\eqlab{PDGbeta}
\eea
\end{subequations}

\begin{figure}[b]
\begin{center}
\includegraphics[width=0.94\linewidth]{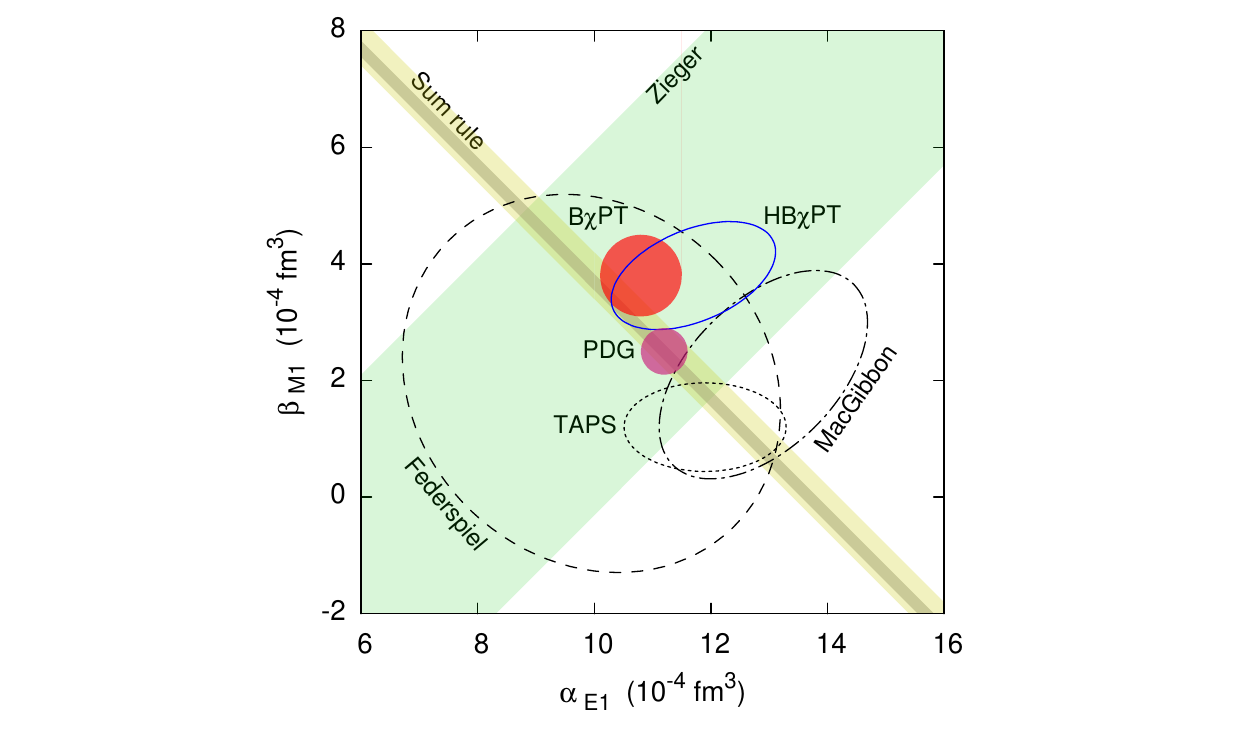}
\caption{ The scalar polarizabilities of the proton. 
Magenta blob represents the PDG summary~\cite{PDG:2013} .
Experimental
results are from Federspiel et~al.~\cite{Federspiel:1991yd},
Zieger et al.~\cite{Zieger:1992jq}, MacGibbon et al.~\cite{MacG95},
and TAPS~\cite{MAMI01}.
`Sum Rule' indicates the Baldin sum rule evaluations of 
$\alpha_{E1}+\beta_{M1}$~\cite{MAMI01} (broader band) and \cite{Bab98}.
ChPT calculations are from \cite{Lensky:2009uv} (B$\chi$PT---red blob)
and the `unconstrained fit' of \cite{McGovern:2012ew} (HB$\chi$PT---blue ellipse).
} 
\figlab{potato}
\end{center}
\end{figure}

These values, together with some other experimental and most recent
theoretical results, are displayed in \Figref{potato}. As the figure shows, the various
determinations of polarizabilities may differ by a few
standard deviations. The main source of these discrepancies is the model
dependence of the extraction of polarizabilities from the unpolarized
Compton scattering cross sections.  The forthcoming measurements
of the beam asymmetry of proton Compton scattering are called for
to sort out this issue \cite{Krupina:2013dya}.

Besides the two scalar polarizabilities, the four spin polarizabilities
of the proton are of significant interest both theoretically and experimentally. 
New experiments at MAMI are aimed to determine them. Two combination of spin polarizabilities have been already determined, 
i.e. forward and backward spin polarizabilities \cite{Drechsel:2002ar}:
\bea
\gamma_0& =& -\gamma_{E1E1}-\gamma_{M1M1}-\gamma_{E1M2}-\gamma_{M1E2}= (-1.0\pm 0.08 \pm 0.1) \times 10^{-4}\, \mathrm{fm}^4,  \\
\gamma_{\pi} &=& -\gamma_{E1E1}+\gamma_{M1M1}-\gamma_{E1M2}+\gamma_{M1E2}= (8.0\pm 1.8 ) \times 10^{-4}\, \mathrm{fm}^4.
\eea
Two more are soon to be measured at MAMI. We shall have a look here at observables
relevant to these measurements.

The polarizabilities arise in the context of low-energy structure of the  nucleon. 
In the process of Compton scattering off the proton $\gamma p \to \gamma p$ they enter as 
coefficients
in the low-energy expansion of the scattering amplitude.
The Feynman diagrams of the process are shown in Fig.~\ref{fig:BornDiags}.
\begin{figure}[t]
\begin{center}
\includegraphics[width=0.7\linewidth]{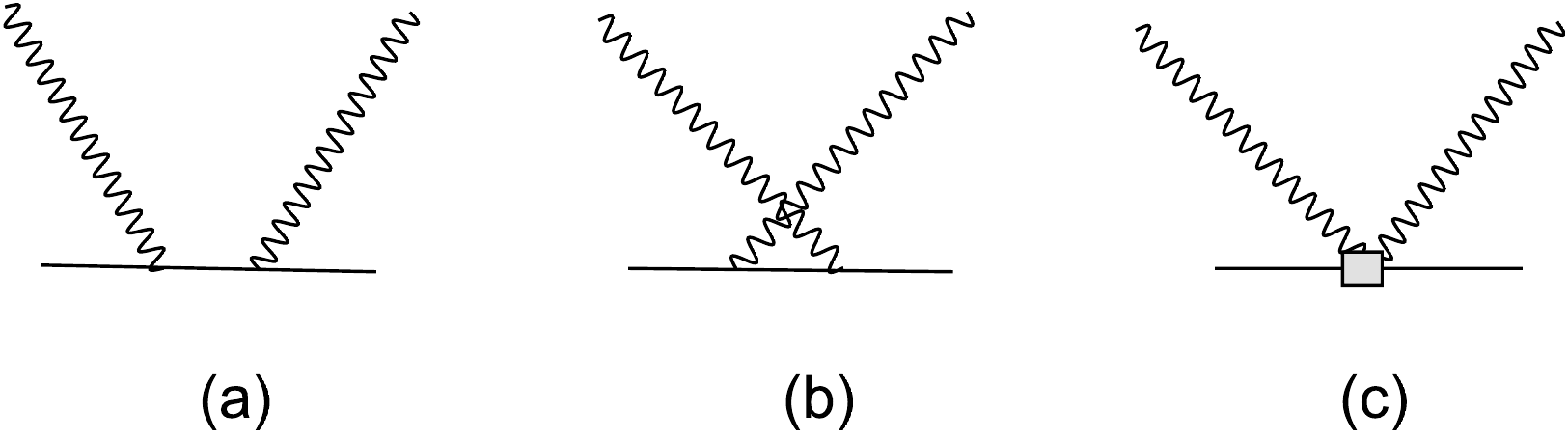}
\caption{Feynman diagrams of low-energy Compton scattering off the nucleon }
\figlab{BornDiags}
\end{center}
\end{figure}
Here graphs~\ref{fig:BornDiags}a and ~\ref{fig:BornDiags}b are the Born contributions,
 assuming that nucleon is a structureless object with mass, electric charge and 
 anomalous magnetic moment. 
% They correspond to the case when the wave length of the photon 
% is much bigger than the size of the proton, so its structure cannot be resolved. As the wave 
%length of 
%the photon decrease, the structure of the photon started to be resolved and this is the place 
% where the polarizabilities starts to contribute. Namely, 
 Graph~\ref{fig:BornDiags}c  is the non Born contribution, and its leading order terms depend on 2 scalar 
 and 4 spin polarizabilities.

Indeed, the scalar polarizabilities starts to contribute at second order in photon energy expansion of the amplitude yielding the following effective Hamiltonian:
\beq
H_{eff}^{(2)} = -4 \pi \Big[ \frac{1}{2} \alpha_{E1} \vec{E}^2 + \frac{1}{2} \beta_{M1} \vec{H}^2 \Big],
\eeq
where  $\vec{E}$ and $\vec{H}$ are the electric and magnetic dipole fields.
 
  In order to introduce scalar polarizabilities in a Lorentz-invariant fashion, we
write down an effective Lagrangian that yields the right
Hamiltonian in the static limit, i.e.,
\bea
\eqlab{effL1}
\lag_{NN\ga\ga} &=& \frac{2\pi }{M^2} (\partial_\alpha \overline N)(\partial_\beta N) (\alpha_{E1}  F^{\alpha \rho} F_{\rho}^{\beta } +\beta_{M1} \widetilde{F}^{\alpha \rho}  \widetilde{F}_{\rho}^{\beta } ) 
\eea
where $F_{\mu\nu} = \pa_{[\mu} A_{\nu]}$ is the electromagnetic field-strength tensor,  $ \widetilde{F}_{\mu\nu} = \epsilon_{\mu \nu \rho \sigma} \partial^{\rho}A^{\sigma}$,
 $N(x)$ is the nucleon Dirac-spinor field.
%  $\eta_{\mu\nu}
%= \mathrm{diag} (1,-1,-1,-1)$ is the Minkowski metric.
Recalling that 
$ 
\widetilde{F}^{\alpha \rho}  \widetilde{F}_{\,\,\,\,\,  \rho}^{\beta } = F^{\alpha \rho} F_{\,\,\,\,\,  \rho}^{\beta }  +\frac{1}{2}  \eta^{\alpha \beta } F^2
 $, 
$
F^2= -2(F^{0i})^2 +  (F^{ij} )^2 =-2 \bE^2 + 2 \bB^2
$,
 and assuming the nucleon rest frame: $\pa_i N=0$,
%$i \ga^0 \pa_0 N = M N $, $i \pa_0 \ol N \ga^0= -M \ol N $,
$ \pa_0 N = -i M N $, $ \pa_0 \ol N = i M \ol N $
we obtain
\bea
\lag_{NN\ga\ga} &= & 2\pi \left\{
\beta_{M1} (\bB^2 -  \bE^2) 
+  (\alpha_{E1}+\beta_{M1})\,\bE^2 \right \}\ol N  N , \nn
\eea
which readily reproduces the well known 
nonrelativistic Hamiltonian: $4 \pi ( 
-\half \alpha_{E1} \bE^2 -\half \beta_{M1} \bB^2) $. 

The Lagrangian
in \Eqref{effL1} can also be rewritten as
\bea
\eqlab{effL2}
\lag_{NN\ga\ga} &=& \pi \beta_{M1} \ol N  N  F^2\, \\
&-& \,
\frac{2\pi(\alpha_{E1}+\beta_{M1} )}{M^2} (\partial_\alpha \overline N)(\partial_\beta N) F^{\alpha \mu} F^{\beta \nu} \eta_{\mu\nu}\nn
\eea

which yields the following Feynman amplitude
\bea
{\cal M}_{\mathrm{NB}}^{\mu\nu} \eps_\mu^\prime \eps_\nu &=&
4\pi \, \ol u(p') \, u(p)  \big[    \beta_{M1} ( q\cdot q' \,  \eps'\cdot \eps
- q\cdot \eps' \, q'\cdot \eps ) 
\nn\\
&-&\frac{\alpha_{E1}+\beta_{M1} }{2M^2}( p_\alpha' p_\beta+ p_\alpha p_\beta')\,  (q^{\prime\alpha} \epsilon^{\prime\mu}
- q^{\prime\mu} \epsilon^{\prime\alpha} ) (  q^\beta \epsilon_{\mu}
- q_\mu \epsilon^{\beta})
\big] \nn\\
&=& \ol u(p') \, u(p) \left[
- A_1^{(\mathrm{NB})}(s,t) \,  \cE' \cdot \cE + 
A_2^{(\mathrm{NB})}(s,t) \,  q\cdot \cE'\, q' \cdot \cE \right],
\eea
where $p$ and $q$ ($p'$ and $q'$) are the four-momenta of incident (outgoing) nucleon and photon, and the manifestly gauge-invariant
polarization vectors are
%\begin{subequations}
\beq
\cE_\mu  = \epsilon_\mu - \frac{(p'+p)\cdot \epsilon}{(p'+p)\cdot q} \, q_\mu\, ,
\quad\quad
\cE_\mu'  = \epsilon_\mu' - \frac{(p'+p)\cdot \epsilon'}{(p'+p)\cdot q} \, q_\mu'\, .
\eqlab{modifiedPolarVector}
\eeq
%\end{subequations}
The polarizability contribution to the invariant 
Compton amplitudes is thus given as follows:
\begin{subequations}
\bea
A_1^{(\mathrm{NB})}(s,t) &=& 2\pi  (\alpha_{E1}+\beta_{M1} )(\nu^2 +\nu^{\prime 2}) 
+ 2\pi  \beta_{M1} t\, ,
\\
A_2^{(\mathrm{NB})}(s,t) &=& -4\pi \beta_{M1} -\pi (\alpha_{E1}+\beta_{M1}) \, t / (2 M^2).
\eea
\eqlab{Acovar}
\end{subequations}

We note that the contribution of $\alpha_{E1}+\beta_{M1}$  differs from conventional
definitions by terms of higher order in the Mandelstam variable $t$, and, hence in energy. For instance, the difference of the present $A_1^{\mathrm{(NB)}}$ with the  
one in Ref.~\cite{Griesshammer:2012we} is equal to $- (\pi/M^2) (\alpha_{E1}+\beta_{M1} )\,  t(\w^2-\quarter t) $.

The last ingredient one needs to obtain the cross section is
the proportionality factor between the matrix element squared and the
cross section:
\beq
\frac{4\pi \al^2 }{(s-M^2)^2} d t.
\eqlab{propfactorCS}
\eeq

This factor can also be expressed in terms of the solid angle $\Omega_L$ by using $d t = (\nu'^2/\pi) d \Omega_L$, where $\nu'$ is the outgoing photon energy.

The previous measurements of the scalar polarizabilities of nucleons were done in unpolarized 
 Compton scattering experiments. 
%Unpolarized nucleon Compton scattering experiments have been done in the past to measure the scalar polarizabilities 
%of the nucleons.  
The non-Born (NB) part
of the unpolarized differential cross section for
Compton scattering off a target with mass $M$ and charge $Ze$
is given by \cite{Baldin}:
\bea
\frac{d\si^{\mathrm{(NB)}} }{d\Omega_L}  &=&-   \frac{Z^2\alpha_{em}}{M} 
\left(\frac{\nu'}{\nu} \right)^2 \nu \nu' \big[ \alpha_{E1} \left( 1 + \cos^2\th_L \right) \nn\\
&+& 2
\beta_{M1} \cos\th_L \big] + O(\nu^4),
\eea
where $\nu = (s-M^2)/2M$ and $\nu' =(-u+M^2)/2M$
are, respectively, the energies of the incident and scattered photon
in the laboratory frame, $\th_L$ ($d\Omega_L=2\pi \sin\th_L d\th_L$) is the scattering (solid) angle;
$s$, $u$, and $t=2M(\nu'-\nu) $ are the Mandelstam variables; and $\al_{em}=e^2/4\pi$ is the fine-structure constant.
Hence, given the exactly known Born contribution \cite{Powell:1949}
and the experimental angular distribution at very low energy,
one could in principle extract the polarizabilities with a negligible
model dependence. In reality, however, in order to resolve
the small polarizability effect in the tiny Compton cross sections,
most of the measurements are done at energies exceeding 100 MeV, 
i.e.,
not small compared to the pion mass $m_\pi$. It is $m_\pi$,
the onset of the pion-production branch cut, that
severely limits the applicability of a polynomial expansion in energy
such as LEX. At the energies around the pion-production
threshold one obtains a very substantial 
sensitivity to polarizabilities but needs to resort to a model-dependent
approach in order to extract them (see~\cite{Drechsel:2002ar,Schumacher:2005an} for reviews).

The magnetic polarizability $\beta_{M1}$ seems to be affected the most:
the central value of the baryon chiral perturbation theory (BChPT) 
calculation is a factor of 1.5 larger than the PDG value. 
This is attributed  to the dominance
 of $\al_{E1}$ in the unpolarized cross section. Thus it is desirable to find an observable
sensitive to $\beta_{M1}$ alone, such that
the latter could be determined independently
of $\al_{E1}$.

Having this in mind, we found that the beam asymmetry could be such an observable. 
It is defined as
\beq
\Sigma_3 \equiv \frac{d\si_{||} - d\si_\perp}{d\si_{||} + d\si_\perp},
\eqlab{BSA}
\eeq
where $d\si_{||} $ and $d\si_\perp$ are
cross sections for photons polarized parallel and perpendicular
to the scattering plane respectively.

Applying the LEX for the beam asymmetry 
 we arrive at the following result for the proton ($Z=1$):
\beq
\label{eq:leadingLEX}
 \Sigma_3 = \Sigma_3^{(\mathrm{B})} - 
\frac{4 M \omega^2 \cos\th \sin^2\th  }{\al_{em} (1+\cos^2\th)^2}\,\beta_{M1} 
+O(\w^4),
\eeq
where $\Sigma_3^{(\mathrm{B})} $ is the exact Born contribution, while
\bea
\omega &=& \frac{s-M^2 + \half t}{\sqrt{4 M^2 - t}}, \quad
\theta = \arccos\left(1 + \frac{t}{2 \omega^2} \right)
\eea 
are the photon energy and
scattering angle in the Breit (brick-wall) reference frame. 
In fact, to this order in the LEX the formula
is valid for $\w$ and $\th$ being the energy and angle in 
the laboratory or center-of-mass frame.

Equation \eref{leadingLEX} shows
that the leading (in LEX) effect of the electric polarizability
cancels out, while the magnetic polarizability
remains. Hence, our first claim is that 
a low-energy measurement of $\Sigma_3$ can in principle 
be used to extract $\beta_{M1}$ independently
of $\al_{E1}$. 

However, the low-energy Compton experiments on the proton are difficult because of small cross sections and overwhelming QED backgrounds.
 Precision measurement only becomes feasible for photon-beam energies
 above 60 MeV and scattering angles greater than 40 degrees. 
 Thus the  experiments at MAMI are being carried
 at photon energies between 80 and 150 MeV. 
Since  at these
 energies the effect of higher-order terms may become substantial one has
to check the applicability of the leading LEX result.
 One way to do that  is to compare 
 the LEX result with the dispersion-relation calculations or calculations based on chiral perturbation theory.

 %\begin{widetext}

 %\end{widetext}
  \begin{figure}[h]
  \begin{center}
\includegraphics[width=0.76\linewidth]{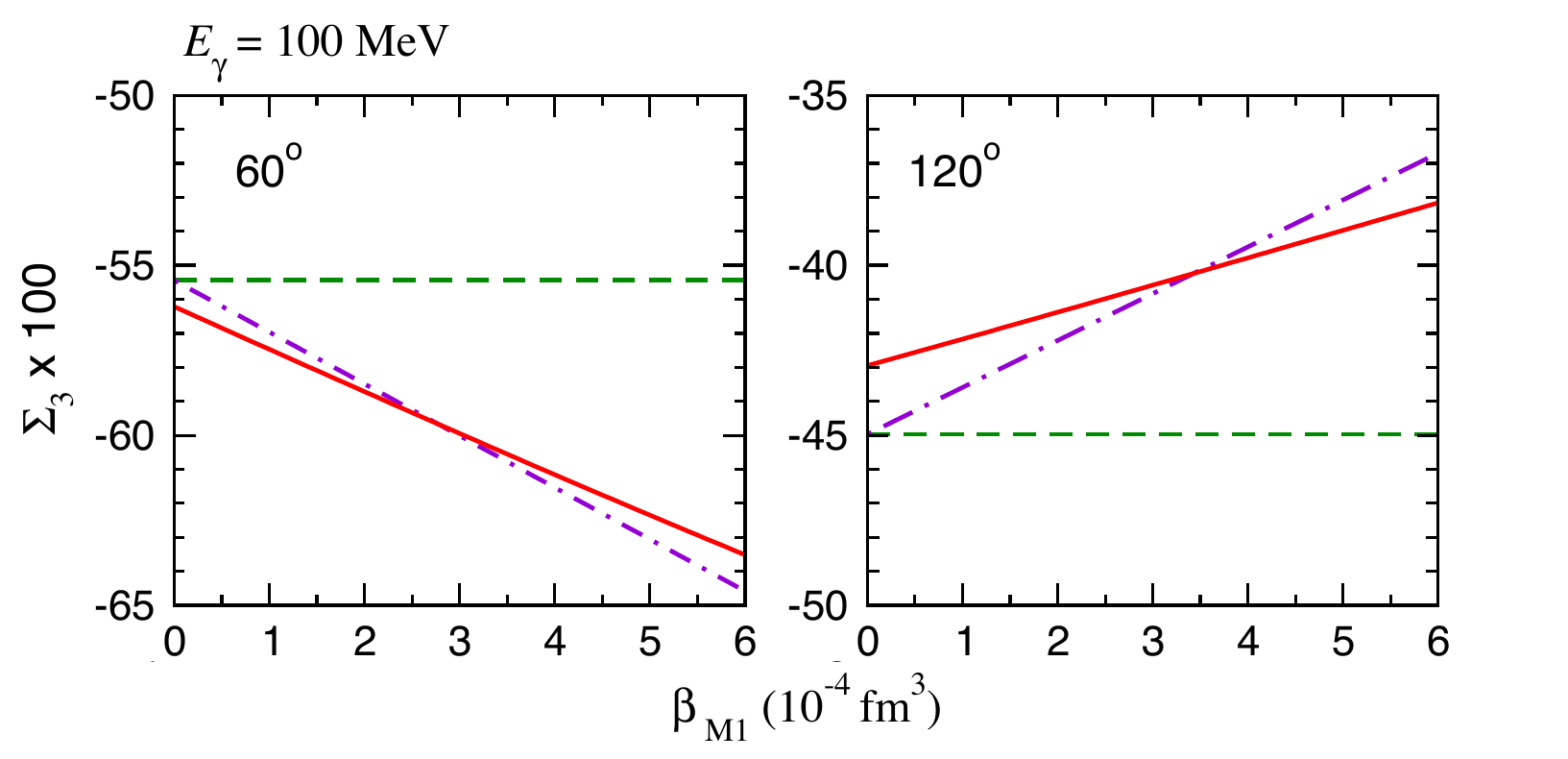}
\caption{Beam asymmetry 
$\Sigma_3$  shown
as function  of $\beta_{M1}$ for fixed photon energy of 100 MeV
and scattering angles of 60 (left panels) and 120 (right panels) degrees.
The curves are as follows: dashed green --- Born contribution; dash-dotted magenta ---
the leading LEX formula Eq. (13);
  red solid --- NNLO BChPT
\cite{Lensky:2009uv}.
}
\figlab{E100}
\end{center}
%\end{minipage}
\end{figure}

\begin{figure}[h]
 \begin{center}
\includegraphics[width=0.76\linewidth]{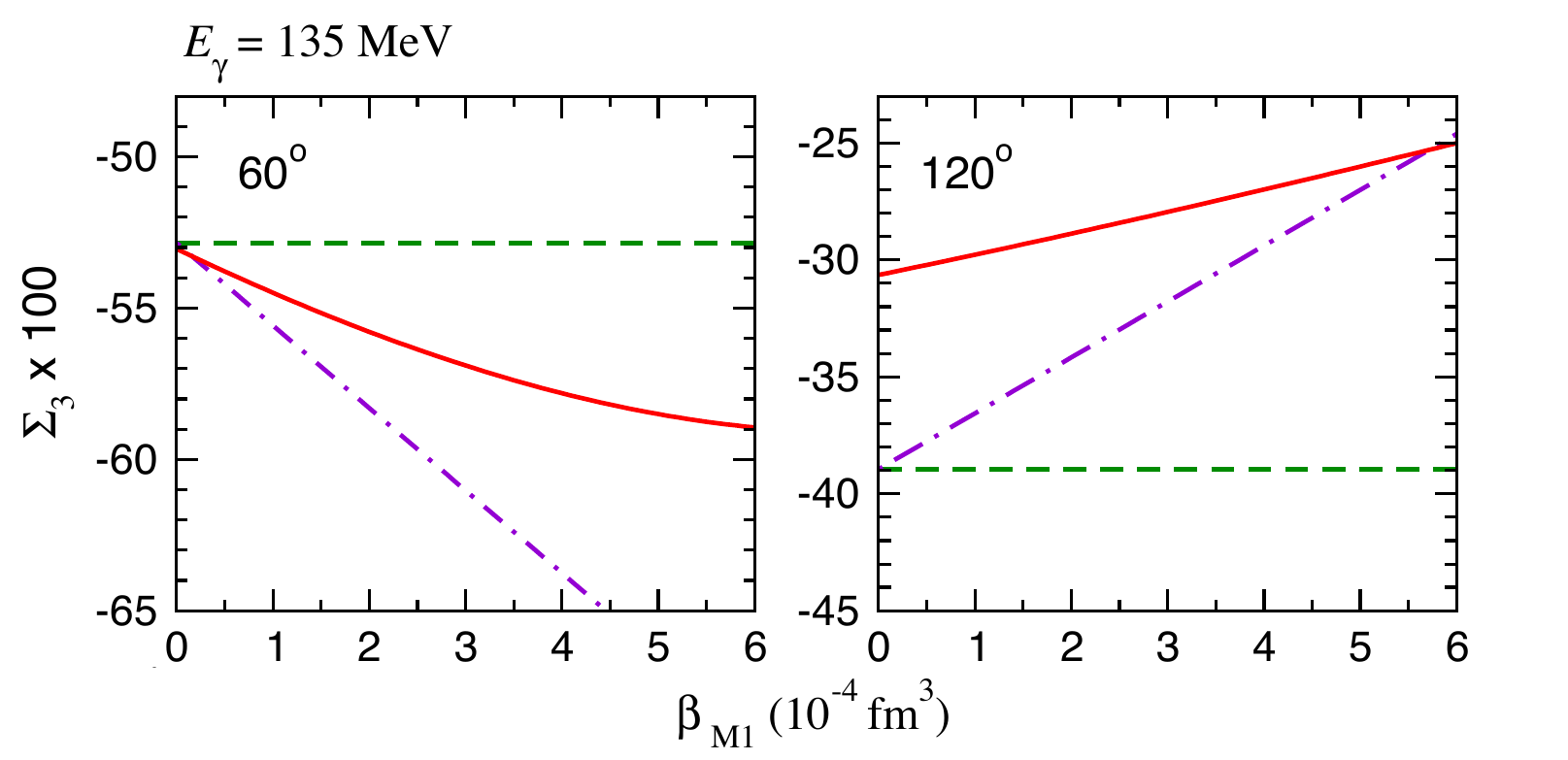}
\caption{The same as in the previous figure but
for photon beam energy of 135 MeV.}
\figlab{E130}
\end{center}
%\end{minipage}
\end{figure}

 Figures~\ref{fig:E100} and \ref{fig:E130} demonstrate such a comparison of
 the leading-LEX result  to the next-next-to-leading order (NNLO) 
 BChPT result of Ref.~\cite{Lensky:2009uv} for the beam asymmetry
 defined in Eq.~\eref{BSA}. The observable
 is plotted for the case of proton Compton scattering as a function of magnetic polarizability of the proton.
 From \Figref{E100} one sees that for the beam energy of 100 MeV
 the LEX is in a good agreement with the BChPT result, especially for the
 forward directions (left panels).

 As expected we observe a significant
 sensitivity of $\Sigma_3$ to $\beta_{M1}$. Also, \Figref{E100} shows that  the beam asymmetry is large and, given the fact that 
many systematic errors tend to cancel out in this observable, the required accuracy to discriminate between the PDG and ChPT values for
the magnetic polarizability should be much easier to achieve. Still,  very high-intensity 
photon beams would be required to achieve the statistics necessary
to pin down the magnetic polarizability
model-independently to the accuracy currently claimed by the PDG, c.f.~\Eqref{PDGbeta}. The high-intensity electron facility MESA 
being constructed in Mainz is very promising in this respect.

 The results for the beam energy of 135 MeV (\Figref{E130})  show that
 the leading LEX result does not apply at such energies. 
% The differences
% between the LEX and BChPT curves at both 100 and 135 MeV 
% are mainly due to $\al_{E1}$ (which here is equal to
% zero for the LEX and about 11 for the BChPT) and the $\pi^0$-anomaly
% contribution.  While it is well known that the anomaly contribution is 
% proportional to $t/(t-m_\pi^2)$ and hence is maximal at the backward angles, 
% the $\al_{E1}$ contribution needs to be examined. 

We next turn to the spin structure of the nucleon. 
It starts to show up at third order in photon energy in the expansion of 
Compton amplitude, yielding the following effective Hamiltonian \cite{Babusci:1998ww}:
\bea
H_{eff}^{(3)} = -4 \pi \Big[ \frac{1}{2} \gamma_{E1E1} \vec{\sigma} \cdot (\vec{E} \times \dot{\vec{E}})+ 
\frac{1}{2} \gamma_{M1M1} \vec{\sigma} \cdot (\vec{H} \times \dot{\vec{H}})
  \\ \nn 
   - \gamma_{M1E2} E_{ij} \sigma_i H_j  + \gamma_{E1M2} H_{ij} \sigma_i E_j\Big],
\eea
here $\dot{\vec{E}} = \partial_t{\vec{E}}$, \, $\dot{\vec{H}} = \partial_t{\vec{H}}$, \, $E_{ij} = \frac{1}{2} (\nabla_i E_j + \nabla_j E_i ) $, \, $H_{ij} = \frac{1}{2} (\nabla_i H_j + \nabla_j H_i ) $. Four constants 
$\gamma_{E1E1}, \,  \gamma_{M1M1}, \, \gamma_{E1M2}$ and $\gamma_{M1E2}$ denote the spin polarizabilities.

Their contribution to the third order can be seen explicitly in  the matrix-element  in the Breit frame, that is given by 

\bea
\eqlab{helAmp}
 T_{\sigma'\lambda' , \sigma \lambda} &=& \be'  \cdot  \be \, A_1(\omega, \theta) +  \be' \cdot \hat{\bq} \,  \be' \cdot  \hat{\bq}' \, A_2(\omega, \theta) \\ \nn
& +& i \bsig \cdot (\be' \times \be) \, A_3(\omega, \theta) + i \bsig \cdot ( \hat{\bq}' \times  \hat{\bq}) \,\be' \cdot \be\, A_4(\omega, \theta) \\ \nn
&+&[ i \bsig \cdot (\be' \times  \hat{\bq}) \, \be \cdot  \hat{\bq}' -   i \bsig \cdot (\be \times  \hat{\bq}') \, \be' \cdot  \hat{\bq} ]\, A_5(\omega, \theta) \\ \nn
& +& [  i \bsig \cdot (\be' \times  \hat{\bq}') \, \be \cdot  \hat{\bq}' -   i \bsig \cdot (\be \times  \hat{\bq}) \, \be' \cdot  \hat{\bq} ]\, A_6(\omega, \theta) 
\eea
where $\bq$ and $\be$ ($\bq'$ and $\be'$)   are momentum and polarization vector of the 
incoming (outgoing) photon, hats indicate unit vectors,  $\omega$ and $\theta$ are its energy and 
scattering angle in the Breit frame; $A_1 - A_6$ functions are invariant amplitudes with the LEX 
expansion given by

\bea
\nn
A_1 &=&-Z^2 +\frac{1}{4}  \left[ (Z+\kappa)^2 (1+z)-Z^2\right] (1-z) \, \omega ^2+ 4 \pi  (\alpha_{E1} +\beta_{M1}  z) \, \omega ^2  + O(\omega ^4), \\ \nn
A_2 &=& \frac{1}{4} \kappa \, z (2 Z + \kappa) \,  \omega ^2 - 4 \pi  \beta_{M1}  \omega ^2 + O(\omega ^4),  \\ \nn
A_3 &=& \frac{1}{2}  \left[ Z (Z+2 \kappa) - (Z+\kappa)^2 \, z \right] \,  \omega - 4 \pi \, \omega ^3
   \left[\gamma _{\text{E1E1}}+\gamma _{\text{E1M2}}+z \left(\gamma
   _{\text{M1E2}}+\gamma _{\text{M1M1}}\right)\right]  + O(\omega ^4),  \\ \nn
A_4 &=& -\frac{1}{2}
    (Z+\kappa)^2 \,  \omega - 4 \pi \,  \omega ^3 \left(\gamma _{\text{M1M1}}-\gamma _{\text{M1E2}}\right)  + O(\omega ^4),  \\ \nn
A_5 &=& \frac{1}{2}   (Z+\kappa)^2 \, \omega + 4 \pi  \, \omega ^3 \, \gamma _{\text{M1M1}} + O(\omega ^4),  \\
A_6 &=&-\frac{1}{2}   Z (Z + \kappa) \, \omega +  4 \pi  \, \omega ^3 \, \gamma _{\text{E1M2}} + O(\omega ^4). 
\eea
Here $z=\cos \th$, $Ze$ and $\kappa$ are the charge and anomalous magnetic moment of the nucleon.

Knowledge of the helicity amplitudes of \Eqref{helAmp} allows one to construct various observables 
and study their sensitivity to spin polarizabilities.
Two observables turn out to be of particular interest for determination of spin polarizabilities: 
the beam target asymmetries with circularly-polarized photons and longitudinally (transversely) 
polarized target, i.e. $\Sigma_{2z}$ ($\Sigma_{2x}$).
Applying the LEX for the beam-target asymmetries, we obtain that the leading  non-Born terms  in 
the Breit frame are:
 \begin{subequations}
\bea
\Sigma_{2x} -   \Sigma_{2x}^{\mathrm{(B)}} &=& \frac{ \sin \theta \, \omega^3}{
   \left(1+z^2 \right) \alpha_{em}} \{
   \alpha_{E1} [(1+\kappa)^2 - (1+2 \kappa) z ]  \nn \\
&+&  \frac{\beta_{M1}}{1+z^2} [\kappa +3 (1+\kappa)^2 z - 3 (1+2 \kappa) z^2 -(1+\kappa)^2 z^3 +(1+\kappa) z^4 ] \\
&+& 2 \left[\gamma _{M1M1} +z \left(\gamma _{E1E1}+\gamma
   _{E1M2}\right) +z^2 \, \gamma
   _{M1E2}\right]\}, \nn \\
\Sigma_{2z} -   \Sigma_{2z}^{\mathrm{(B)}} & =&  \frac{ \, \omega^3}{ (1+z^2) \alpha_{em} }  \{ \alpha_{E1} [-\kappa + 2 (1+\kappa)^2 z - (2+3 \kappa) z^2]  \nn \\
&+& \frac{\beta_{M1}}{1+z^2} [-(1+\kappa)^2+(1-\kappa ) z +6 (1+\kappa)^2
   z^2 -2 (3+4 \kappa) z^3 -(1+\kappa)^2 z^4+(1+\kappa) z^5] \nn \\
&+& 2 [(1+z^2) ( \gamma _{E1E1} + z  \, \gamma _{M1E2} ) + 2 z (\gamma _{M1M1}+z \,  \gamma _{E1M2}) ] \}.
\eea
  \eqlab{Sigma2}
 \end{subequations}
Unfortunately, the applicability of \Eqref{Sigma2} is very limited. 
 \begin{figure}
 \begin{center}
\includegraphics[width=0.76\linewidth]{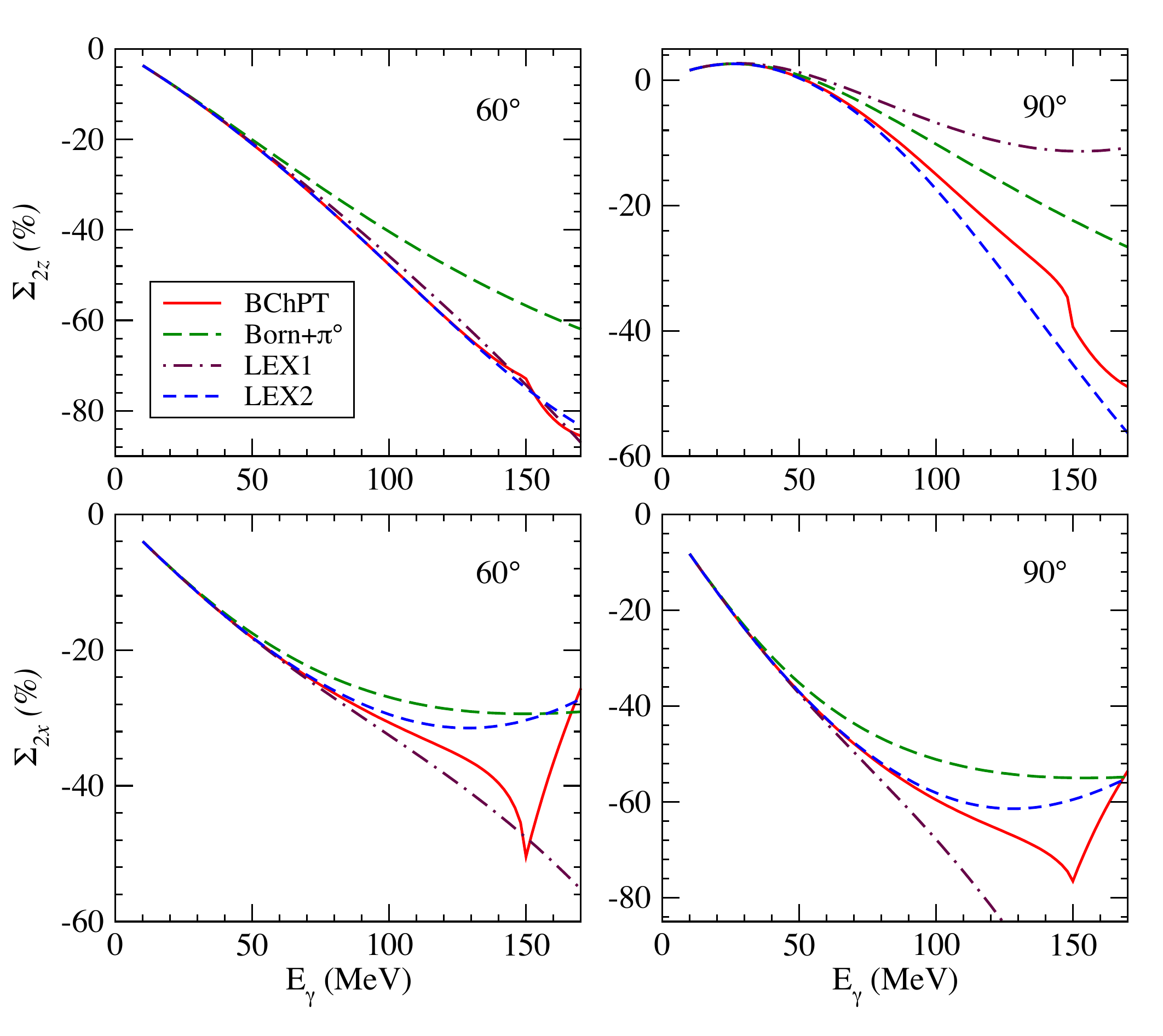}
\caption{ The beam-target asymmetries $\Sigma_{2z}$ (upper panel) and $\Sigma_{2x}$ (lower 
panel)  as a function of incident photon energy for scattering angle of  60  (left panel) and
  90  (right panel) degrees.  The curves are as follows: dashed green --- Born contribution; 
   red solid --- NNLO BChPT;  dashed blue --- the  LEX with only  invariant amplitudes expanded; 
   dash-dotted magenta --- the leading LEX formulas  Eqs. (18)
   with both the invariant 
  amplitudes and helicity amplitudes expanded. }
\figlab{Edep_Sigma2}
\end{center}
\end{figure}
Similarly to the case of the LEX for the beam asymmetry, we define it by comparing
 the leading order LEX results of \Eqref{Sigma2} with results obtained in BChPT. 
  Figure~\ref{fig:Edep_Sigma2}  demonstrates such a comparison.
 Two LEX curves correspond to the expansion of the invariant amplitudes 
 (LEX2 curve) and additionally the expansion of the helicity amplitudes 
 (LEX1 curve given by \Eqref{Sigma2}). One sees that LEX and 
 BChPT curves coincide only for photon energy below 50 MeV, 
 thereby defining the region of applicability.
However, at these low energies (below 50 MeV), one finds the leading order LEX 
in \Eqref{Sigma2} to be suppressed by  $\omega^3$. The sensitivity 
to spin polarizabilities becomes too small
to allow one to extract them from current experiments.
Therefore, the spin polarizabilities are planned to be extracted
at higher energies (around  the Delta resonance region), 
where the sensitivity of the observables  becomes significant. As discussed above, the LEX approach 
fails, and one has to resort to either dispersion relations or ChPT approach. 

To conclude, we claim that
the beam asymmetry
$\Sigma_3$ should be used for accurate determination 
of the magnetic polarizability $\beta_{M1}$ from low-energy Compton scattering. 
While the cross sections  receive contributions from both the
electric  and magnetic  polarizability, the effect of $\al_{E1}$
cancels out from the asymmetry at leading order in the low-energy expansion.
We have also studied the next-to-leading corrections  and found them to be suppressed at the
forward scattering angles \cite{Krupina:2013dya}.  A precise 
and model-independent determination of
the proton $\beta_{M1}$ is feasible through
a precision measurement of $\Sigma_3$ at beam energies
below 100 MeV and forward scattering angles.
Furthermore,  when multiplied with the unpolarized cross section, $\Sigma_3$
yields the polarized cross section difference, which
provides an exclusive access to the electric polarizability.

Besides the scalar polarizabilities,  we have studied observables
 sensitive 
to spin polarizabilities. 
The problem here is the small region of applicability of the LEX results,
 i.e. photon energy below 50 MeV. At such energies the  sensitivity of LEX results to spin
  polarizabilities becomes
too small to discriminate their effect from the Born contribution.
In this case, one has to resort to either ChPT or dispersion relation approaches. Both work at higher 
energy regimes, and the fact that sensitivity  to spin polarizabilities increases increasing the energy, 
suggests an idea to extract them around the Delta resonance region, where the sensitivity  becomes 
substantial.
Nevertheless, we obtained the LEX expressions for the non-Born leading order terms of beam-target 
asymmetries $\Sigma_{2z}$ and   $\Sigma_{2x}$, cf.~\Eqref{Sigma2}. 
Although one cannot use these expressions for determination of the spin polarizabilities,
they could provide a low energy test for either the ChPT or dispersion relation frameworks.

\medskip
%\section*{Acknowledgements}
I would like to thank Vladimir Pascalutsa for advising me during this work, and to acknowledge the support of the  Graduate School DFG/GRK 1581
``Symmetry Breaking in Fundamental Interactions".


\begin{thebibliography}{99}

\bibitem{Griesshammer:2012we} 
  H.~W.~Griesshammer, J.~A.~McGovern, D.~R.~Phillips and G.~Feldman,
  %``Using effective field theory to analyse low-energy Compton scattering data from protons and light nuclei,''
  Prog.\ Part.\ Nucl.\ Phys.\  {\bf 67}, 841 (2012).
 % [arXiv:1203.6834 [nucl-th]]. 
 
 %\cite{Holstein:2013kia}
\bibitem{Holstein:2013kia}
  B.~R.~Holstein and S.~Scherer,
  %``Hadron Polarizabilities,''
  arXiv:1401.0140 [hep-ph].
  %%CITATION = ARXIV:1401.0140;%%
  %1 citations counted in INSPIRE as of 02 May 2014


%\cite{Beringer:1900zz}
\bibitem{Beringer:1900zz} 
  J.~Beringer {\it et al.}  [Particle Data Group Collaboration],
  %``Review of Particle Physics (RPP),''
  Phys.\ Rev.\ D {\bf 86}, 010001 (2012).
  %%CITATION = PHRVA,D86,010001;%%

  

 
%\cite{PDG:2013}
\bibitem{PDG:2013}   
  J.~Beringer {\it et al.}  [Particle Data Group Collaboration], updated online edition (2013)
  http://pdg.lbl.gov/2013/tables/rpp2013-sum-baryons.pdf

%\cite{Krupina:2013dya}
\bibitem{Krupina:2013dya}
  N.~Krupina and V.~Pascalutsa,
  %``Separation of proton polarizabilities with the beam asymmetry of Compton scattering,''
  Phys.\ Rev.\ Lett.\  {\bf 110} (2013) 26,  262001
  [arXiv:1304.7404 [nucl-th]].
  %%CITATION = ARXIV:1304.7404;%%
  %5 citations counted in INSPIRE as of 30 Apr 2014

%\cite{Drechsel:2002ar}
\bibitem{Drechsel:2002ar}
  D.~Drechsel, B.~Pasquini and M.~Vanderhaeghen,
  %``Dispersion relations in real and virtual Compton scattering,''
  Phys.\ Rept.\  {\bf 378}, 99 (2003).
  %[arXiv:hep-ph/0212124].
  %%CITATION = PRPLC,378,99;%%
  
  %\cite{Babusci:1998ww}
\bibitem{Babusci:1998ww}
  D.~Babusci, G.~Giordano, A.~I.~L'vov, G.~Matone and A.~M.~Nathan,
  %``Low-energy Compton scattering of polarized photons on polarized nucleons,''
  Phys.\ Rev.\ C {\bf 58} (1998) 1013
  [hep-ph/9803347].
  %%CITATION = HEP-PH/9803347;%%
  %97 citations counted in INSPIRE as of 02 May 2014
  
  \bibitem{Baldin}
  A.~M.~Baldin,
  %``Polarizability Of Nucleons,''
  Nucl.\ Phys.\  {\bf 18}, 310 (1960).
  
  %\cite{Powell:1949}
\bibitem{Powell:1949} 
 J.~L.~Powell,  Phys.\ Rev.\  {\bf 75}, 32 (1949).
 
 \bibitem{Schumacher:2005an}
  M.~Schumacher,
  %``Polarizability of the nucleon and Compton scattering,''
  Prog.\ Part.\ Nucl.\ Phys.\  {\bf 55}, 567 (2005).
  %[arXiv:hep-ph/0501167].
  %%CITATION = PPNPD,55,567;%%

  %\cite{Lensky:2009uv}
\bibitem{Lensky:2009uv} 
  V.~Lensky and V.~Pascalutsa,
  %``Predictive powers of chiral perturbation theory in Compton scattering off protons,''
  Eur.\ Phys.\ J.\ C {\bf 65}, 195 (2010).
 % [arXiv:0907.0451 [hep-ph]].
  %%CITATION = ARXIV:0907.0451;%%
  
 
  \bibitem{Federspiel:1991yd}
  F.~J.~Federspiel {\it et al.},
  %``The Proton Compton effect: A Measurement of the electric and magnetic
  %polarizabilities of the proton,''
  Phys.\ Rev.\ Lett.\  {\bf 67}, 1511 (1991).
  %%CITATION = PRLTA,67,1511;%%

\bibitem{Zieger:1992jq}
  A.~Zieger, R.~Van de Vyver, D.~Christmann, A.~De Graeve, C.~Van den Abeele and B.~Ziegler,
  %``180-degrees Compton scattering by the proton below the pion threshold,''
  Phys.\ Lett.\  B {\bf 278}, 34 (1992).
  %%CITATION = PHLTA,B278,34;%%

\bibitem{MacG95}
B.~E.~MacGibbon, G.~Garino, M.~A.~Lucas, A.M.~Nathan, G.~Feldman and B.~Dolbilkin,
%``Measurement of the electric and magnetic polarizabilities of the proton,''
Phys.\ Rev.\ C {\bf 52}, 2097 (1995).
%[arXiv:nucl-ex/9507001].
%%CITATION = NUCL-EX 9507001;%%

\bibitem{MAMI01}
V.~Olmos de Leon {\it et al.},
%``Low-Energy Compton Scattering And The Polarizabilities Of The Proton,''
Eur.\ Phys.\ J.\ A {\bf 10}, 207 (2001).
%%CITATION = EPHJA,A10,207;%%

\bibitem{Bab98}
D. Babusci, G. Giordano and G. Matone,
Phys. Rev. C {\bf 57}, 291 (1998).

  \bibitem{McGovern:2012ew} 
  J.~A.~McGovern, D.~R.~Phillips and H.~W.~Griesshammer,
  %``Compton scattering from the proton in an effective field theory with explicit Delta degrees of freedom,''
  Eur.\ Phys.\ J.\ A {\bf 49}, 12 (2013).
%  [arXiv:1210.4104 [nucl-th]].


\end{thebibliography}
\end{document}